# Photonic Pseudo-Random Number Generator for Internet-of-Things Authentication using a Waveguide based Physical Unclonable Function

CHARIS MESARITAKIS,[1] PANAGIOTIS RIZOMILIOTIS,[2] MARIALENA AKRIOTOU,[3] CHARIDIMOS CHAINTOUTIS,[4] ALEXANDROS FRAGKOS,[4] AND DIMITRIS SYVRIDIS[,3]

[1]Dept. Information & Communication Systems Engineering University of the Aegean, 2 Palama str., Karlovassi, Samos, GR-83200, Greece
[2]Dept. Informatics and Telematics, Harokopio University of Athens, 9, Omirou str., 17778, Athens, Greece
[3]Dept. Informatics & Telecommunications, National & Kapodistrian University of Athens, Panepistimiolois Ilisia, 15784, Greece
[4] Eulambia Advanced Technologies Ltd. Ag. Ioannou 24, 15342, Athens, Greece

**Abstract:** In this paper we experimentally evaluate a physical unclonable function based on a polymer optical waveguide, as a time-invariant, replication-resilient, source of entropy. The elevated physical unclonability of our implementation is combined with spatial light modulation and post processing techniques, thus allowing the deterministic generation of an exponentially large pool of unpredictable responses. The quality of the generated numbers is validated through NIST/DIEHARD(ER) suites, whereas the overall security of the scheme is benchmarked assuming attackers with elevated privileges in terms of system access. Finally, based on the demonstrated key features, we present and analyze a mutual authentication implementation scenario which is fully compatible with state-of-the-art commercial Internet-Of-Things architectures.

## 1. Introduction

The internet-of-things (IoT) expanding ecosystem has triggered new requirements regarding the physical and cyber security of smart devices. In particular, the majority of implementations consists of highly inter-connected devices deployed in physically unsupervised premises or of devices with minimum or low-quality security features, due to cost considerations [1]. In this heterogeneous and volatile landscape, Physical Unclonable Functions (PUFs) start emerging as physical roots of trust [2] able to provide replication resilient hardware modules that could be employed either as authentication tokens [3,4] or as counterparts to software based pseudo-random generators [5–8]. In particular, PUFs are physical systems, whose transfer function is too complex to be reliably evaluated or reproduced, thus enabling a one-way transformation of the input (challenge) to a unique output (response). Despite, the associated physical complexity, the challenge-response mapping is a deterministic process and therefore, if the same PUF structure-input pair is employed, then the same response will be obtained. This pivotal feature if combined with the ability to generate an exponential number of uncorrelated challenge-response pairs and an enhanced resiliency to noise, can enable PUF modules to act as clone-resilient hardware pseudo-random number generators. In particular, by exploiting strong PUF modalities [9] the generated numbers-keys will not have to be stored in non-volatile memory in the devices but can be reproduced on-demand. Thus, the cyber-security features of the overall system can be significantly hardened. Towards this direction multiple PUF implementations have been proposed, spanning from typical silicon-microcircuit PUFs [10–13] to optical modalities [14–18]. Although silicon-cast PUFs provide the best footprint-cost performance, they can be vulnerable to physical cloning, simulation or entropy reduction through side channel-attacks; In this landscape, optical PUFs, despite their increased footprint, emerge as an attractive solution that has proven its hardness against physical cloning [19,20].

The generic principle of operation of optical PUFs, relies on optical scattering that occurs when a coherent source illuminates a semi-transparent material, that has a large number of scattering centers. The interaction of the phase and amplitude component of scattered photons at an imaging plane away from the PUF's volume, allows the generation of spatially complex images (speckle), whose features are directly related to the intensity-phase- spatial distribution of the utilized wavefront, the laser's wavelength and the exact location of scatters in the PUF's volume [21]. Based on this principle, optical PUFs have exploited the use of spatial light modulators (SLM) so as to modulate the spatial profile of the radiation that illuminates their surface, allowing the deterministic generation of pseudo-random responses (speckles) [21,22]. Nonetheless, in all these systems the number of discrete challenges escalates linearly with the number of projected pixels of the SLM at the PUF's volume. This drawback stems from the fact that optical scattering, at least when employing moderate optical power levels, is a linear process [21,22]. Therefore, by using combinations of pixels at the SLM, the generated speckle patterns will be a linear combinations of the patterns generated from each individual pixel. This is a key problem because the number of challenge-response pairs sets a hard threshold regarding security against brute-force attacks, whereas linearity renders the system vulnerable to machine-learning attack scenarios [22].

Recently, we demonstrated an optical PUF based on a large-core optical waveguide [20]. The proposed system demonstrated orders of magnitude increased resiliency to physical cloning compared to conventional implementations, whereas it was set to operate as a weak PUF, meaning that security was guaranteed only by the physical uniqueness of the PUF and the "randomness" of its responses under different challenges was not evaluated. In this work, we expand our initial concept of a waveguide-based-PUF, so as to generate an exponentially large pool of responses and validate their efficiency in terms of their statistical similarity to true random generators (TRNGs). Contrary to previous implementations, the response's linearity is bypassed through the utilization of post-processing techniques and maximum distance coding (MDC) hashing, that exploit the spatial complexity of the generated speckle patterns. Evaluation of the experimental set of challenge-response pairs was performed through the help of standard suits like NIST/DIEHARD(ER) and by tools like context-weight-tree (CTW) compression. Furthermore, contrary to [20], we have revamped the proposed PUF device so as to explicitly consists of low-cost and readily available parts so as to minimize cost and footprint. The hardware requirement reduction renders our scheme more suitable for IoT architectures; thus, we describe an implementation scenario of an optical PUF hardened mutual-authentication protocol, which is compatible with state-of-the-art IoT deployments like Amazon's Greengrass [23] and Google's IoT Cloud [24].

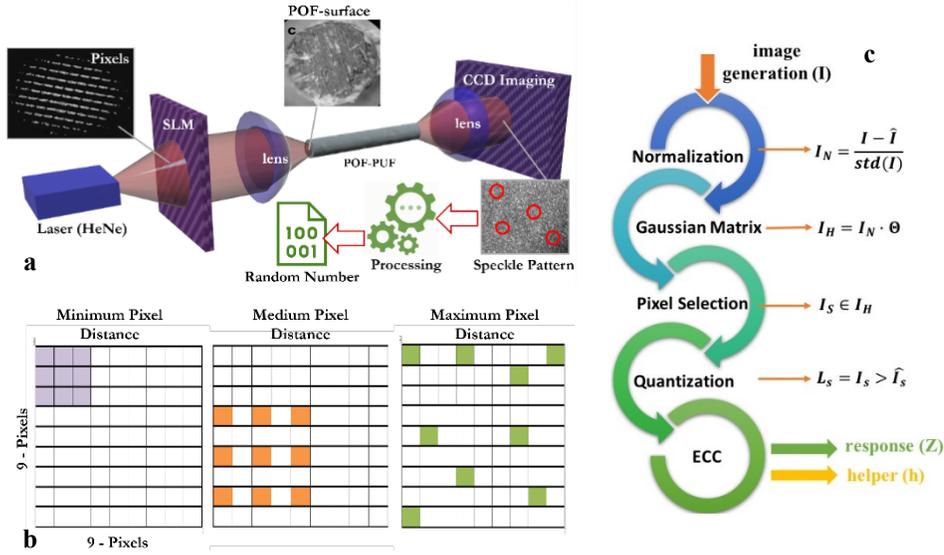

Fig. 1.a Schematic of the basic experimental setup. SLM stands for spatial light modulator whereas POF for polymer optical fiber. Insets depict from left to right: image of the SLM pixels illuminated by the laser source, microscope image of the POF's surface and a typical speckle pattern. b. Three different challenge generating patterns at the SLM. c schematic representation of the Random Binary Hashing Technique

## 2. Experimental Setup & Raw Response Processing

The experimental setup is depicted in Fig.1a; it incorporates a single mode HeNe laser tube emitting at 632.8nm with output power of 1mW as the coherent optical source. A 240x340 liquid-crystal-display (LCD) has been placed prior to the laser acting as an SLM; each pixel could be switched either on or off, thus allowing $2^N$ possible spatial states where N corresponds to the number of pixels projected to the POF's facet, through an objective lens. The projecting optics were set so as to enable 9x9 SLM's pixels to be imaged to the PUF's surface (0.785mm$^2$) without evident diffraction patterns (inset of fig.1a). The PUF-core consists of a commercial polymer optical fiber (POF), with 980 μm core, 20 μm cladding and reduced length compared to [20] equal to 3cm. The boundary conditions of the fiber allow in principle the guiding of more than $2 \cdot 10^6$ discrete transverse modes. The facets of the POF were processed through a randomly driven friction system aiming to induce random defects at the surface (inset of fig.1a). As analyzed in [20] these defects, alongside in-fiber impurities, mechanical bends etc. enable the randomization of the inter-modal power distribution. Taking into consideration that these modes exhibit different phase velocity and spatial profile, when they superimpose at an imaging plane (Fraunhofer diffraction) away from the POF's output, allow the generation of a highly complex speckle-like pattern. The projection of the transverse modes is achieved through a second objective lens, whereas the imaging plane consists of a typical charge-couple device (CCD) imaging module with 1280x960 pixels (inset fig.1a). The SLM and CCD are simultaneously controlled/monitored by a low cost micro-controller.

The aforementioned setup enables the generation of an exponential number of challenge-response pairs ($2^{81}$). In particular, through the SLM's pixels we can regulate the spatial profile of the beam that illuminates the POF's facet, thus we can vary the initial transverse mode power distribution. Taking into consideration that the limited fiber length hinders modal power equilibrium [25], each spatial pattern at the input leads to a different mode-mixture at the output and thus to discrete speckle patters at the imaging plane. Following this methodology, we have experimentally acquired three discrete challenge-response sets, that correspond to different SLM's pixel sets (fig.1b). Each set consisted of 3x3 pixels, whose combinations allowed the extraction of $2^9$ challenge-response pairs and thus a total of 1536 challenges out of the complete pool of $2^{81}$ combinations. The first set uses adjacent pixels (fig.1b) aiming to provide minimum spatial variation among potential challenges and thus benchmark the system under stringent conditions, while the second and third sets comprise pixels that have larger spatial separation, thus offering lower spatial overlap.

The recorded responses are fed into and processed by a personal computer so as to enable random number generation. The post processing procedure is based on the Random Binary method and is described in detail in [19,20]. Briefly, the hashing procedure is depicted in fig.1c; it consists of intensity normalization for each image and multiplication with a constant random matrix (Θ), whose values exhibit normal distribution. It is worth mentioning that this matrix is fixed for all images thus does not contribute to the "randomness" of the response. The final stage includes the random selection of pixels, following a uniform distribution ($I_s$). The results are quantized based on their mean value and a binary string is extracted from each image (Z). Finally, a typical error correcting code (ECC) like BCH or equivalent is included so as to amend the detrimental effects of noise in the system (helper data - h). Similar to previous works [19,20], all metrics regarding the unpredictability, unclonability or robustness of PUF's results, are evaluated versus the number of redundant ECC bits.

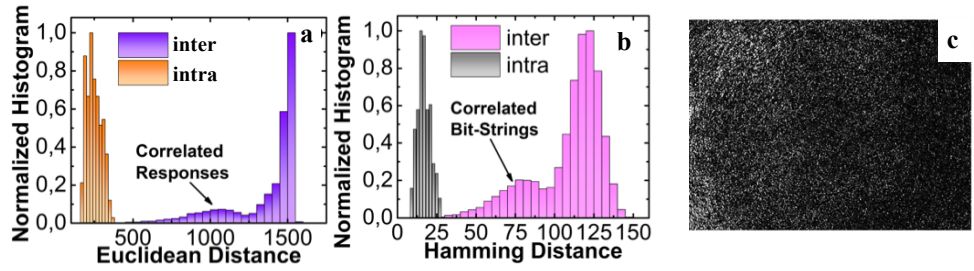

Fig. 2. Euclidean distance for the two discrete data sets, intra corresponds to 60 responses from the same PUF instance and same challenge, while inter originates from the same PUF but using combinations of the SLM's pixels as different challenges. b. hamming distances for the same data-sets as in a. c) typical raw response recorded at the CCD,

## 3. Experimental Results & Analysis

3.1 Photonic PUF pseudo-random number generator

The scenario assumed in this work comprises employing the PUF module as a deterministic optical pseudo-random number generator; meaning that responses under different challenges should be uncorrelated and thus unpredictable, whereas the overall procedure should be noise resilient. From an operational point of view, each PUF's challenge can be considered to provide a similar functionality to that of a seed in conventional algorithmic generators. Contrary to previous works, as challenges all SLM's pixel combinations are employed and not only pixel primitives [22]. Consequently, the generated responses (raw speckle images) on one hand form an exponentially large pool but at the same time, they are expected to be linearly dependent. The first metric employed is the Euclidean distance for two discrete data sets; the first includes all the responses (images) for all the challenges (inter-distance Fig.2a), while the second reflects the system's resiliency to noise and includes 60 responses for the same challenge at the SLM (intra-distance Fig.2a) over a span of several minutes. Inter-distance exhibits a mean value of 245 that originates from pixel intensity variations, while intra-distance is considerable higher with a mean value of 1474. The existence of an overlap for these two distributions, corresponds to an increased probability for false-positive/negative. In our case, although the two distributions marginally do not overlap, it can be observed that the inter-distance distribution exhibits an elongated tail. This stems from the linear nature of the optical scattering process involved in response generation. In particular, this effect is manifested when pairs of challenges are used that that have the same SLM pixel in an "ON" state. For example, in a two pixel case ($P_1$ and $P_2$); the correlation of responses that correspond to challenge ($P_1$:ON, $P_2$:OFF) or ($P_1$:OFF, $P_2$:ON) and ($P_1$:ON, $P_2$:ON) is significantly higher compared to ($P_1$:ON, $P_2$:OFF) and ($P_1$:OFF, $P_2$:ON). It is obvious that for a scenario involving multiple pixels ($P_{1...N}$) the correlation of ($P_1$:ON, $P_{2...N}$:OFF) and ($P_{1...N}$:ON) would be significantly lower compared to the case mentioned above. This is the reason that the majority of inter-distance samples provide high Euclidean distance (fig.2a).

A potential argument to the effect shown above is that this statistical "anomaly" can be eradicated through the image hashing technique applied to the raw responses. In order to validate this hypothesis, we generated a set of 256-bit long strings by hashing each response with the process depicted in fig.1c. It is worth mentioning that the same random Gaussian matrix and the same pixel selection matrix was used for all responses ($\Theta$-$I_s$). In fig.2b the hamming distance (number of bit-flips) have been computed for the hashed versions of all responses, originating from both data sets (inter & intra). It is clear that the elongated tail persists, even after hashing, meaning that the random binary procedure projects pixel intensities to binary strings but does not contribute to the inherent randomness of the raw response. In detail, the mean hamming distance of the intra case is computed to be 17.8 implying a noise floor of 6.9% bit flips, while the inter case's mean is 120.3 leading to a 46.9% bit flips. This deviation from the ideal randomness of 50% is assumed to originate from the cumulative effect of linear correlation among images and to the existence of constant spatial features, due to ambient light during experimental measurements.

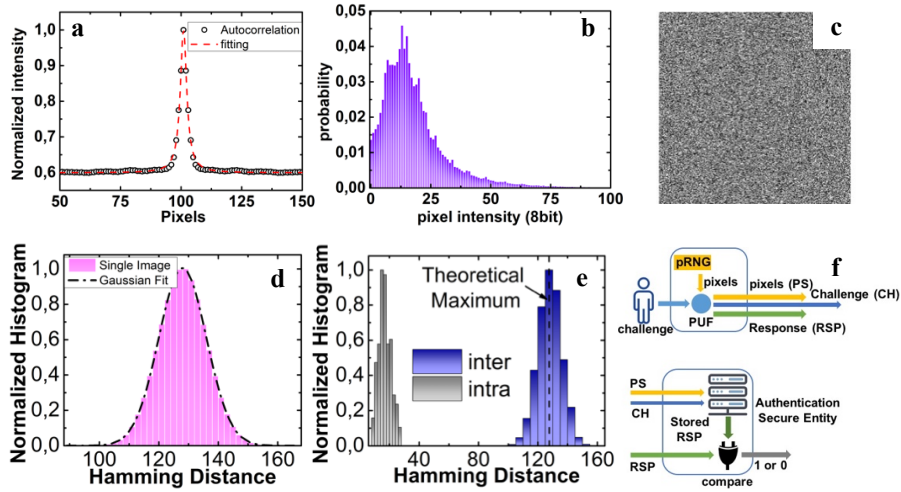

Fig. 3. Euclidean distance for the two discrete data sets, intra corresponds to 60 responses from the same PUF instance and same challenge, while inter originates from the same PUF but using combinations of the SLM's pixels as different challenges. b. hamming distances for the same data-sets as in a. c) typical raw response recorded at the CCD,

3.2 Enhancing Unpredictability through Spatial Complexity

Based on the aforementioned results, the statistical properties of the generated responses provide significant bias (fixed bit values). This is the reason why previous works on optical PUFs, apart from using a different physical medium, avoided the combination of pixels at the challenge generating element (SLM). On the other hand, if only primitive pixels are employed the number of potential responses scales linearly with the number of pixels and consequently is not adequate for real-life key generation applications. Aiming to circumvent this problem, we devised an alternative strategy. In fig.2c a characteristic raw response (speckle pattern) is shown, exhibiting a massive number of spatial features (light-dark spots). It is intuitive that although each image possibly possesses a significant amount of entropy we extract only a tiny fraction (256bits) from this available pool. Aiming to quantify the entropy of each image and thus compute the number of random bits (N) present in each response, we use the following equation $N = \frac{X \cdot Y}{d} \sum_{i=1}^{256} I_i \cdot log_2(I_i)$, where d is the mean speckle size, computed through estimating the full width at half maximum of the autocorrelation of the image, X and Y are the vertical and horizontal pixels in the image and $I_i$ is the probability for each intensity level from an available 256 states (8bit depth). As shown in Fig.3a the image autocorrelation after fitting, assuming Lorentzian distribution, provided a mean speckle size d=3.2 pixels that is equal to 1.6 times the Nyquist criterion. In Fig.3b the intensity histogram for a typical response is demonstrated, following a gamma function as dictated by typical speckle theory for unpolarised speckles [26]. Based on these computations, a typical response from our system can provide ≈2·10$^6$ random bits, thus up to now, in each challenge-response pair we extracted only ≈0.012% of the available randomness.

Following this lead, we preserve the same random Gaussian matrix during hashing (see fig.1e) and we apply the same challenge at SLM, while we vary the selection matrix ($I_s$) so as to extract multiple bit strings even from a single image. Thus, we constructed 100 different bit-strings from a single response and in Fig.3c we have plotted them (logical '1' and '0' is depicted as white/black respectively); it can be seen that visual inspection provide no evident spatial pattern. In fig.3d we have computed the hamming distances of these bit strings and the mean value is at the theoretical limit (50%) whereas the histogram follows a perfect Gaussian distribution. Therefore, if we adapt the challenge generating mechanism and we assume that for each LCD pixel combination, a random set of 256 pixels at the CCD will be also chosen, then the generating responses will proliferate from the entropy present at each image. In this case, the challenge will consist of a vector for the LCD pixels and another for the CCD pixels (pixel selection in Fig.1e), while the response will be again a 256bit long string. In Fig.3e we have recomputed the hamming distances for the dataset used in fig2a-b but using the aforementioned two-factor challenge strategy (pixels selection and challenge). It is worth mentioning that this alternative technique has no effect on the hamming distance of the intra-data set, due to the fact that noise is equally present at all CCD pixels independently of the selection process. On the other hand, the inter-distance has changed radically, the elongated tail of the distribution has vanished, the mean value is set to the ideal 50% whereas the distribution also follows a perfect Gaussian shape. Therefore, the lack of any overlap between the two distributions (intra/inter) nullifies the probability of false positive/negative.

Nonetheless, the proposed solution has two flaws; the first is the fact that our photonic PUF aims to replace conventional pseudo-random generators, while through this approach it relies on their existence and their quality so as to provide secure operation (pixel selection). In this context, the dependence of the PUF to pseudo-random generator support renders the system vulnerable to side-channel attacks, while optical PUFs typically does not suffer from such a drawback. The second problem arises when the device is employed in an authentication scenario. In such a case (fig.3f) the user challenges the PUF so as to extract a binary response (RSP) and transmits it alongside the challenge (CH) [19]. A secure entity has stored the responses, during a system enrollment phase, and uses the CH so as to extract its own (RSP)', whereas by comparing RSP⊗RSP' authentication is achieved. In this modified scenario, the PUF should also transmit the pixel selection (PS) to the secure entity (fig.3f). The overhead imposed by such an approach scales with the length of the generated random number. For example, a key 256bit long, demands at least 4096 additional bits per RSP (256·16bit for CCD's pixel coordinates).

3.3  Spatial Complexity and Maximum Distance Coding (MDC)

Aiming to preserve the enhanced entropy of each image but eliminate the flaws introduced by the use of a pseudo-random generator, we propose an optimized scenario for challenge generation. In particular, instead of randomly constructing the pixel selection vector for each challenge, we feed the binary representation of the challenge (sequence number representing each pixel at the SLM) to the BCH encoder used for error correction. The computed code-words are constructed so as to exhibit maximum hamming distance. The length of the MDC-assisted challenge is dictated by Galois fields [27] and can be chosen so as to be larger than $K \cdot 2^m$, where K is the random number bit length and m≥$\log_2$(X·Y), with X, Y being the dimensions of the image in pixels. Taking into consideration that the binary input in the BCH are spatial coordinates, maximum distance coding at the challenge results in selecting pixels that exhibit maximum spatial distance among them. More importantly, the pixel selection can be precomputed and can be public, thus there is no need for transmitting an extensive challenge, contrary to the scenario depicted in Fig.3f. Furthermore, this approach can provide a "non-random" way to construct the pixel selection vector for each challenge and thus eliminates the need for an algorithmic pseudo-random generator alongside the vulnerabilities that such an addon imposes.

Following this approach, each challenge is projected through the BCH to a fixed set of pixel coordinates that are used during the hashing; we have used this methodology for the same data sets as the ones used in Fig.2b and Fig.3e. In fig.4a the hamming distance for the inter-class is presented; it is obvious that the mean hamming distance is set to the theoretical maximum and no statistical anomalies can be detected. Obviously, we do not plot the intra-class distribution because similarly to fig.3e the alternative CCD pixel selection process does not affect the noise level and thus bit-flips. Taking a step further, we used the security framework used in [19,20] but now instead of computing the probability of physical cloning, we estimate the probability of a duplicate RSP for different CH (see fig. 4b). An important part of the framework is that each RSP is accompanied by helper data (not the CH and PS) that include redundant bits produced again through a BCH encoder, so at to shield against noise induced bit-flips. This step is critical if our system is used as a deterministic source of randomness, meaning that the user instead of storing his/her seed/key reproduces it on demand by reapplying the same CH to the same PUF. Under this assumption the RSP should be time invariant and thus we include the error correcting code (ECC). Taking into consideration that ECC can reduce the variation due to noise but also affect the RSP changes that occur due to CH variation, the probability of prediction is computed versus the number of ECC bits similarly to [19, 20]. In fig.4b the probability that two different CH correspond to the same RSP versus the ECC bits is computed for the standard hashing procedure (see section 3.1) and for the modified procedure that corresponds to changing the PS during each CH, as mentioned in this section. It can be seen that the security offered by the waveguide PUF through the modified PS and CH procedure is significantly strengthened. In particular, with the inclusion of 55 ECC bits the typical PUF processing offers a probability of prediction of $10^{-2}$, while with through the modified approach the probability scales down by 5 orders of magnitude to $10^{-7}$. It is worth mentioning that for fewer ECC bits the probability with the proposed technique is zero.

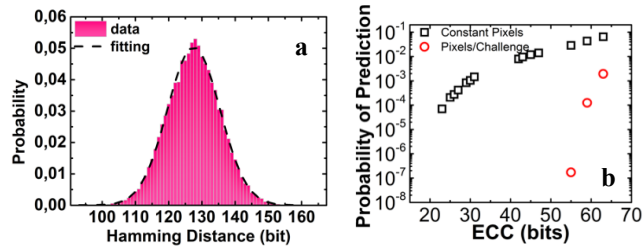

Fig. 4. Hamming distance for the inter-dataset and for 256bit long bit strings. b) probability of prediction versus the number of ECC for the simple scenario with constant PS and for the optimized case of MDC assisted PS

3.4 Randomness evaluation of the MDC-assisted challenge generation

The aforementioned results provide insight that the waveguide-based optical PUF discussed here, apart from being replication resilient [20], can be also used as a deterministic photonic pseudo-random number generator. Nonetheless, in order to validate such a claim, we tested the PUF-generated bit strings using standard randomness tests. The first test is the NIST suite that evaluates the similarity of the statistical properties of the target generator to a true random source [28]. The basic requirements imposed by the NIST suite so as to provide meaningful results, is to apply the test on a data set of at least 1Gbit for a level of certainty α-0.1. In order to fulfil this requirement, we exploited the total entropy per response. From an initial set of approximately 1536 images we hashed each response multiple times, by changing in each case the PS, so as to extract 4000 256bit long responses per image. $4000 \times 256 = 1$Mbit, which is smaller than the maximum 2Mbit random bits present in each image as computed in section 3.2. The results for the 15 NIST tests are provided in table 1. The proposed waveguide PUF generator passes all the tests, while preserving uniformity (p-value), therefore the generated responses cannot be distinguished statistically from a true-random source. It is worth mentioning that the tests marked with an asterisk correspond to multiple tests and only the worst value is present at the table I.

In the same context, the aforementioned dataset has been fed to the DIEHARD suite, that is also a collection of tests that benchmarks the efficiency of the generator as a random source [29]. For comparative reasons DIEHARD suite has been fed also with a second dataset originating from a true random generator (TRNG). In table 2 the results (p-values) for 17 tests are presented for the proposed PUF whereas for the TRNG only a pass-fail indication is provided. It is worth mentioning that the PUF generated results passed all the tests from this suite as well, whereas interestingly the TRNG provided some "weak" results (marginal success) for multiple tests, whereas PUF driven numbers provided only a single "weak" result. Aiming to use a stringer test we also used the DIEHARDER suite which is a set of additional 14 tests. In this case the optical PUF responses failed in four tests, while TRNG failed in five tests and provided multiple "weak success" results. These results validate the statistical similarity of the generated responses to TRNGs. A final test used by the cryptographic community is the resiliency of the generated bit-strings to data compression techniques [30]. One of the key algorithms used for compression of binary data is the context-weight-tree algorithm [31]. For benchmarking we employed PUF's data, a regular text file (CIA fact book), a file containing only logical '0', bit strings generated by pseudo-random algorithm [32] and the TRNG source used above. So as to allow comparison all data sets had the same size. In table 3 the compression rate (Cr) is presented for each technique in bits/byte. If $C_r=8$, means that no compression was achieved, while if $C_r<<8$ means that significant amount of redundancy was evident at each dataset's byte. It is evident that the lack of any entropy in the file containing only zeros results to high compression (Cr=0.005bits/byte), whereas a typical English text provided also a high compression ratio. On the other hand, the Ziggurat algorithm, the PUF and the TRNG provided similar results (Cr=8bits/byte) that dictate high resiliency to compression due to high entropy. This means that CTW also confirms that from a statistical point of view the optimized PUF module provide indistinguishable results to a TRNG.

**Table 1. NIST Test PUF Results**

| Test | p-Value | Success (%) |
| --- | --- | --- |

| Test | p-Value | % |
|---|---|---|
| Block Frequency | 0.7645 | 98.75 |
| Cumulative Sums-1 | 0.4895 | 98.5 |
| Entropy | 0.6111 | 99% |
| FFT | 0.3306 | 98.5 |
| Mono-Bit | 0.9512 | 98.75 |
| Linear Complexity | 0.7349 | 98.75 |
| Longest Run | 0.689 | 97.75 |
| Non-Overlapping Templates-1 | 0.1372 | 97.75 |
| Overlapping Templates | 0.7498 | 98.25 |
| Random Excursion Variable-1 | 0.3431 | 98.18 |
| Random Excursions | 0.1394 | 97.27 |
| Rank | 0.1969 | 99.75 |
| Runs | 0.2429 | 98.75 |
| Serial | 0.1473 | 98 |
| Universal | 0.0308 | 99.5 |

Table 2. DIEHARD(ER) PUF/TRNG Test Results

| Test | p-Value (No.) | PUF | TRNG |
|---|---|---|---|
| Birthday Test | 0.2351 | pass | pass |
| Operm5 | 0.000005 | weak | pass |
| Rank 32x32 | 0.9407 | pass | pass |
| Rank 6x8 | 0.8511 | pass | pass |
| Bitstream | 0.6166 (20) | pass | weak |
| Opso | 0.9291 (23) | pass | pass |
| Oqso | 0.6205 (28) | pass | pass |
| Dna | 0.5877 (31) | pass | pass |
| Count 1s str | 0.3582 | pass | pass |
| Count 1s byte | 0.6851 | pass | pass |
| Parking Lot | 0.4982 | pass | pass |
| 2D Spheres | 0.4488 | pass | pass |
| 3D Spheres | 0.9504 | pass | pass |
| Squeeze | 0.267 | pass | weak |
| Sums | 0.0535 | pass | pass |
| Runs | 0.8361 | pass | pass |
| Craps | 0.368 | pass | pass |
| **Dieharder** | | | |
| Tsang gcd | 0 | fail | fail |
| Monobit | 0.5442 | pass | pass |
| Runs sts | 0.9448 | pass | weak |
| Serial* | 0.1288 (31) | pass | pass |
| RGB bit dist.-1 | 0.0766 (12) | pass | pass |
| RGB min Dist.-1 | 0.0288 (5) | pass | weak |
| RGB perm.-1 | 0.1761 (4) | pass | pass |
| RGB lagged -2 | - (32) | pass/fail | pass/fail |
| RGB Kstest | 0.454 | pass | pass |
| Byte distr. | 0 | fail | fail |
| DCT | 0.9235 | pass | pass |
| Filltree-1 | 0.0586 (2) | pass | weak |
| Filltree 2-1 | 0.000584 | weak | fail |
| Monobit2 | 1 | fail | fail |

Table 3. CTW Compression Results

| Source | Cr |
|---|---|
| Zero Padding File | 0.00586 |
| CIA Fact book | 1.44178 |
| Ziggurat | 8.00755 |
| PUF | 8.00752 |
| TRNG | 8.00754 |

## 4. Mutual authentication using photonic PUF as root for trust

In this section, we combine the merits of the proposed scheme with the IoT paradigm. We assume two mutual authentication scenarios where a single base station (BS) with internet access is set to authenticate multiple users/devices. The BS either serves as the intermediate between the sensors and a cloud infrastructure or it is the main computation resource in the case of edge devices. We assume that the IoT devices that are used can perform public key operations and that the SSL/TLS 1.2 (or 1.3) is supported. This is the most common use case that we encounter in practice. The authentication is divided into two phases. In the first phase, the Handshake protocol of SSL/TLS is used to authenticate the BS. The BS possesses a private/public key pair that it is used to prove its identity over a public channel. When this phase is completed successfully, a secure channel is established of the between the BS and the device. In the second phase, the device is authenticated using the secure channel and the SSL/TLS Record protocol. We distinguish two cases depending on the type of credentials that are available, the public key and hybrid use cases.

In the public key use case, the device, also, has a private/public key pair. The public key is disseminated using a X.509 certificate and the private must be stored in the device. The certificate can be either sent when the device tries to connect or can be stored at the BS when the device's owner registers the device. The device proves its identity by signing, with its private key, a short message. In most of the cases this is a JSON Web token (JWT). Finally, the BS verifies the signature and the identity with the public key from the certificate (fig.5a). In the hybrid scenario, symmetric key encryption is used for the authentication of the device. This approach is similar to password-based authentication where the device and the BS share the same secret key. At some point, the device was registered at the BS and this secret key was securely stored in a local database of the BS. The same secret key is stored also at the device. After the authentication of the BS, the SSL/TLS Record protocol protects the communication between the two entities. Thus, the device can securely send the secret key unencrypted via the secure channel and prove its identity (fig.5b).

The BS and the IoT devices must store and protect different secret keys depending on the authentication scenario (Table 4a). In the public key use case, the BS and the device must only store its own private key. In the hybrid scenario, the BS must store its own private key as well as all the secret keys of the registered devices, while the IoT device must store the secret key that it shares with the BS. All these keys must be securely. Instead of using dedicated and expensive secure hardware (usually FIPS certified), we propose a PUF-based alternative. More precisely, instead of storing the keys, we compute them on-the-fly as the output of the PUF.

|  | Public key Scenario | Hybrid Scenario |
|---|---|---|
| Base Station | 1Private key | 1Private key |
|  |  | N secret keys |
| IoT Device | 1Private key | 1secret key |

Table 4. Number of keys stored in each player for two scenarios

In more detail, all the public key schemes have a set-up phase, in which, the private/public key pair is computed. In the vast majority of the cases, like in the DSA digital signature algorithm, the private part of the key pair is a randomly selected value R and, then, the public part of the key is deterministically computed from the private key. The PUF can be easily integrated in the key pair generation process. That is that, instead of using the randomly selected value R directly as the private key, the random value R can be first filtered by the PUF and, then, use the output as the private key. Thus, each time the private key is needed, R is fed in the PUF and the private key is produced on-the-fly and it is never stored. The confidentiality of R doesn't have to be protected (Fig. 6). Similarly, in the symmetric key, the secure storage of the shared secret can be replaced by the on-the-fly computation of the key each time the key is needed. The key is again the output of PUF, using a random value R as input, and, once more, the confidentiality of R doesn't have to be protected.

However, the unclonability property of the PUF imposes a restriction. The BS and the IoT device cannot both use the "same" PUF to store the shared key. One of the two entities must use secure hardware. Thus, the hybrid authentication scenario has two variants. In the first case, the PUF is used by the IoT devices, and the BS uses secure memory hardware to store all the shared keys (Fig. 7.a). In the second case, the PUF is used by the BS and each device must possess secure memory hardware. In that case, the BS only has to store unprotected the random values R that corresponds to each device (Fig. 7.b). Table 2 summarizes the design choices.

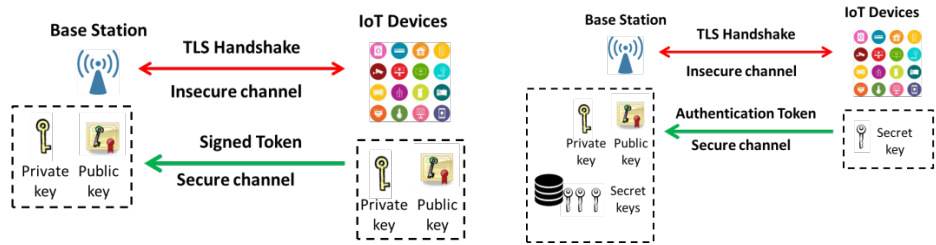

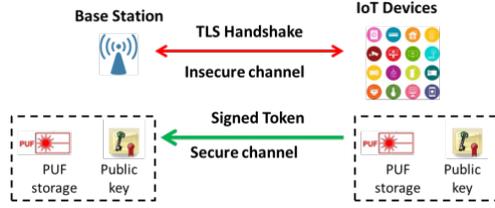

Fig. 5.a Schematic of the public key based authentication scenario. b. Schematic of the hybrid (public key and symmetric key) authentication scenario.

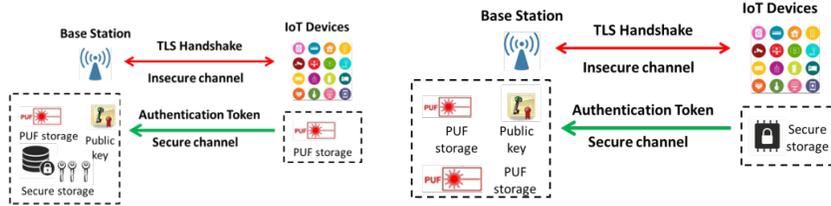

Fig. 6 Schematic of the public key based authentication scenario with PUF storage protection.

Fig. 7. a. Schematic of the hybrid authentication scenario with PUF storage protection at the device. b. Schematic of the hybrid authentication scenario with PUF storage protection at the BS.

|  | Public key Scenario | Hybrid Scenario | |
|---|---|---|---|
| **Base Station** | 1 PUF | 1 PUF | 1 PUF |
|  |  | 1 PUF | N secHW |
| **IoT Device** | 1 PUF | 1secHW | 1 PUF |

## 5. Conclusion

In this work we expand our concept of an optical waveguide based PUF so as to generate a low-cost photonic module that can be used as a nonreplicable secure deterministic pseudo-random generator, with the ability to produce an exponentially large pool of responses. Towards this direction, we employed off-the-selves components and reduced the footprint of the experimental prototype to handheld dimensions. The linearity of the optical scattering process that stems when pixel combinations at the challenge creating SLM is used and causes highly correlated responses is bypassed by devising a new response processing procedure. We exploit MDC so as to sample the PUF's response, thus choose pixels that exhibit maximum spatial distance and are linked on the applied challenge. This stratagem is computationally secure, challenge can be precomputed and thus be public. More importantly the generated binary responses were evaluated through standard randomness tests (DIEHARD-ER, NIST, CTW etc.) and provided no statistical deviations from a TRNG.

Finally, the confirmed security of the generated responses, the demonstrated replication resiliency of the module alongside the deterministic process responsible for PUF's response alleviates the need for high-cost and side-channel attack prone procedure of storing cryptographic keys in IoT devices or base-stations; and can allow on-demand reproduction of cryptographic keys. This feature can significantly cyber harden the state-of-the-art IoT architectures.